# High resolution nanofocus X-ray source based on ultracold electrons from laser cooled-atoms

Gholamreza Shayeganrad

**Summary:** X-ray 3D tomography is so far limited to micron resolution because of the large focusing spot of the focused electron beam on the metal target. Here, we proposed, to the best knowledge of the author, for the first time a novel nanofocus X-ray 3D tomography system based on focused ultracold electrons from laser-cooled atoms in a nanoscale region of metal target in vacuum which is useful for 3D nanotomography and submicron volumetric imaging.

**Keywords:** Ultracold electron, X-ray 3D tomography, Cold atom, Photoionization, High resolution.

## 1. Introduction

Computed tomography (CT), initially used in biology [1, 2], and today is also used in materials research in 3D microtomography [3-5]. The technology has been improved by implementing the microfocus X-ray sources (MFX) in the computed tomography facilities. The microfocus X-ray tube is one of the key components of Micro-CT devices because the spatial resolution of the CT image is, in theory, mainly determined by the size of the X-ray focal spot of the X-ray source [6, 7]. In principle, the image spatial resolution is primarily determined by the X-ray source focal spot size, the detector element size and the stability and the mechanical vibrations, whereas the contrast resolution is generaly set by the X-ray flux and also detector element size. In microfocus X-ray computed tomography the resolution of a system decreases (magnitude larger) almost linearly with object size. In Table 1 some specifications of a typical MFX source (L12531) is summarized.

So far, state-of-the-art X-ray tomography is limited to X-ray tube with micron resolution. The proposed sub-microfocus X-ray is an X-ray source based on a cold-electrons from near threshold photoionized laser-cooled atoms that generates X-ray by focusing electrons through an electron lens to sub-micron area on a metal target in vacuum and then X-ray emits from the sub-micron focal spot on the metal target as shown schematically in Fig. 1. The smaller the focal spot, the clear and sharper the acquired X-ray image even with magnified geometrically. The image data can be recorded by a digital detector with high small pixel elements. In other words, the accuracy depends on the focus spot size of X-ray source which is used to generate X-rays, the smaller the X-ray focal spot, the higher the resolution will be.

It is noticeable that ultracold (T<1 K) neutral plasmas has been experimentally realized by photoionizing of ultracold atoms [8-11].

**Table 1.** Specifications of a typical MFX source (L12531).

| Parameters | Specification | Unit |
|---|---|---|
| X-ray tube current operation setting | 10 to 200 | μA |
| Maximum output | 16 | W |
| Maximum resolution | 2 | μm |
| FOD (focus of object distance) | Approx. 1 | mm |

## 2. Method and Materials

For details please refer to Ref. [12]. Briefly, an atom beam, i.e. Cs beam, leaves a pyramidal MOT in an ultra-high-vacuum (UHV) chamber with a residual pressure of approximately $10^{-9}$ mbar. The pyramidal MOT compared to other methods enables a simplified optical setup.

Laser cooling and trapping of atoms is achieved by shining a single laser radiation in the presence of the nonhomogeneous static magnetic field produced by a pair of coils in the anti-Helmholtz configuration. In typical operating conditions, the flux of cold atoms is approximately $5 \times 10^8$ atoms/s, ultimately limited by the instability comes from thermal moving atoms.

The cooled Cs atoms in ground-state are photoionization in near threshold. Afterward electrons are focused by electromagnetic lenses onto a nano-focal spot which is a characteristic of nano-spot focusability of ultracold electrons.

Frequency stability of the excitation laser is maintained by locking it to the atomic line through the saturated absorption-spectroscopy technique. While, ionization can obtained with a tunable laser (for example, R6G or Ti:Sapphire laser).

## 3. Discussion

The core of this setup is using cold electrons from laser cooled atoms. Compared with the X-ray synchrotron radiation emitted from high energy electrons in storage ring facilities to obtain computed tomograms, the proposed system has an advantage of low emittance and therefore better image resolution. In addition, femtosecond X-ray sources based on this technique can be realized with the cost of less electron per pulse. In addition to the mentioned above, it can be used in spatial-time-resolved X-ray spectroscopy in semiconductor and material sciences by using a femto- or pico-second laser for ionization of cold atoms to generate femto- or pico-pulsed cold electrons and consequently ultrafast X-ray pulsed sources. X-ray synchrotron radiation is a high flux of parallel beam radiation, however, the lack of our proposed system can be addressed by increasing integration time and sub-micron focusing spot size. The proposed method permits the use of detectors with very high detection efficiency. Hereby, low power nanofocus X-ray sources compare favorably to synchrotron sources for nanotomography applications.

## 4. Conclusion

In conclusion, a novel, non-destructive, high-resolution tool for quantitative 3D imaging of the medical or industrial materials has been proposed based on high-resolution sub-microfocus X-ray computed tomography (Nano-CT) images. It allows to non-destructively assess not only the roughness of the materials but also inside the structure.

The novel proposed nanofocused X-ray sources compared to the conventional microfocus X-ray tube-based sources have higher resolution. This will extend X-ray inspection applications to non-destructive X-ray Nano CT. Better resolution to sub-micron or nan-range by applying a unique laser cooled electron beam could enhance imaging resolution in biotechnology.

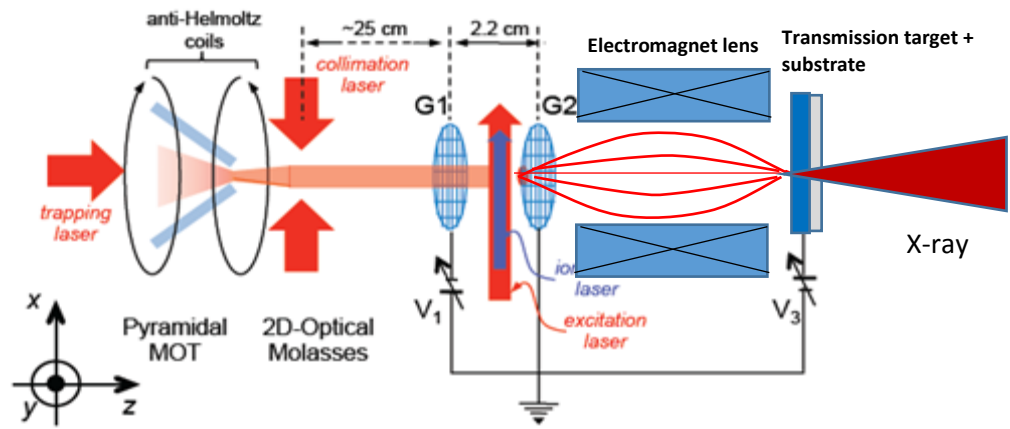

**Fig. 1.** Schematic diagram of proposed nanofocus X-ray source with using cold electrons form laser-cooled atoms.